\documentclass[a4paper,twocolumn,longbibliography,accepted=2026-01-14]{quantumarticle}
\pdfoutput=1

\usepackage{graphicx}
\usepackage{dcolumn}
\usepackage{bm}
\usepackage{color}
\usepackage{amsmath,amsthm,amsfonts,amssymb,bm}
\usepackage{dsfont}
\usepackage[normalem]{ulem}
\usepackage[hidelinks]{hyperref}
\usepackage[table,xcdraw]{xcolor}
\usepackage[utf8]{inputenc}
\usepackage[T1]{fontenc}
\usepackage{physics}
\usepackage{mathtools}
\usepackage{url}
\hypersetup{
    colorlinks,
    linkcolor={blue!80!black},
    citecolor={blue!80!black},
    urlcolor={blue!80!black}
}

\begin{document}

\title{Recurrence in discrete-time quantum stochastic walks}
\author{M. \v{S}tefa\v{n}\'ak}
\affiliation{Department of Physics, Faculty of Nuclear Sciences and Physical Engineering, Czech Technical University in Prague, B\v{r}ehov\'a 7, 115 19 Praha 1-Star\'e M\v{e}sto, Czech Republic}
\author{V. Poto\v{c}ek}
\affiliation{Department of Physics, Faculty of Nuclear Sciences and Physical Engineering, Czech Technical University in Prague, B\v{r}ehov\'a 7, 115 19 Praha 1-Star\'e M\v{e}sto, Czech Republic}
\author{\.{I}. Yal\c{c}{\i}nkaya}
\affiliation{Department of Physics, Faculty of Nuclear Sciences and Physical Engineering, Czech Technical University in Prague, B\v{r}ehov\'a 7, 115 19 Praha 1-Star\'e M\v{e}sto, Czech Republic}
\author{A. G\'abris}
\affiliation{Department of Physics, Faculty of Nuclear Sciences and Physical Engineering, Czech Technical University in Prague, B\v{r}ehov\'a 7, 115 19 Praha 1-Star\'e M\v{e}sto, Czech Republic}
\affiliation{Institute for Solid State Physics and Optics, HUN-REN Wigner Research Centre for Physics, 1525 P.O. Box 49, Hungary}
\author{I. Jex}
\affiliation{Department of Physics, Faculty of Nuclear Sciences and Physical Engineering, Czech Technical University in Prague, B\v{r}ehov\'a 7, 115 19 Praha 1-Star\'e M\v{e}sto, Czech Republic}

\begin{abstract}
Interplay between quantum interference and classical randomness can enhance performance of various quantum information tasks. In the present paper we analyze recurrence phenomena in the discrete-time quantum stochastic walk  on a line, which is a quantum stochastic process that interpolates between quantum and classical walk dynamics. Surprisingly, we find that introducing classical randomness can reduce the recurrence probability --- despite the fact that the classical random walk returns with certainty --- and we identify the conditions under which this intriguing phenomenon occurs. Numerical evaluation of the first-return generating function allows us to investigate the asymptotics of the return probability as the step number approaches infinity. This provides strong evidence that the suppression of recurrence probability is not a transient effect but a robust feature of the underlying quantum-classical interplay in the asymptotic limit. Our results show that for certain tasks discrete-time quantum stochastic walks outperform both classical random walks and unitary quantum walks.
\end{abstract}

\maketitle

\section{\label{sec:intro}Introduction}

Decoherence, disorder, and randomness are generally considered detrimental to quantum information processing. However, several recent studies showed that stochastic randomness can provide advantage in certain tasks, e.g. hybrid \cite{morley_hybrid_2019} and randomized algorithms \cite{wallman_noise_2016, wang_qubit-efficient_2023, wan_randomized_2022, proctor_scalable_2022}, error mitigation \cite{li_efficient_2017, temme_error_2017, kim_scalable_2023}, Hamiltonian simulations \cite{campbell_random_2019}, and quantum walks \cite{Kendon_decoherence_2003,apers_unified_2021}. Some of these applications include randomness as a necessity to mitigate the errors present in NISQ devices, while others consider randomness as a tool to improve performance. The results of these latter studies raise an intriguing question: How universal is this behavior? Can we expect that there exists a broad class of tasks where additional randomness may yield useful advantages? Our work is a step in the direction of trying to answer the latter question, by focusing on quantum walks.

Quantum walks (QW) and classical random walks (RW) exhibit several differences, making QWs an attractive paradigm for exploiting the quantum advantage in computational tasks. Both discrete-time~\cite{Aharonov1993, Meyer1996} and continuous-time~\cite{Farhi1998} QWs have found various applications, such as quantum Monte Carlo~\cite{Montanaro2015} and optimization problems~\cite{Marsh2020, Slate2021, Casares2022}, which generally offer a quadratic speed-up over classical methods~\cite{Melnikov2019} as first pointed out in the context of spatial search on grids~\cite{Aaronson2003,Childs2004,oriekhov_constant_2024} and hypercube~\cite{Shenvi2003, Potocek2009}. More recently, quadratic speed-up over a classical RW search for any number of marked vertices was achieved in both discrete-time~\cite{Ambainis2020} and continuous-time~\cite{Apers2022} QW algorithms.

Quantum stochastic walks (QSW) were originally introduced as a generalization of unitary QWs, allowing dynamics on a graph governed by a quantum stochastic equation of motion~\cite{Whitfield2010}. This theoretical framework enables interpolation between a unitary QW and a classical RW. In this respect, QSWs have been particularly studied in a continuous-time setting, where the evolution is described by the Lindblad master equation. This includes investigations into the decoherence in continuous-time QWs~\cite{Bressanini2022} and the quantum-to-classical transition with stochastic resetting~\cite{Wald2021}. On the application side, QSWs have been explored in various contexts, such as graph isomorphism testing~\cite{Bruderer2016}, quantum state discrimination~\cite{Pozza2020}, data classification~\cite{Wang2022}, quantum reinforcement learning for maze escape~\cite{Caruso2016,Pozza2022}, quantum transport \cite{dudhe_testing_2022,nzongani_nonunitary_2025}, quantum generalizations of PageRank algorithm \cite{garnerone_thermodynamic_2012,benjamin_resolving_2024} or dynamical phase transitions \cite{longhi_dynamical_2025}. It has been established that incorporating stochasticity into unitary quantum evolution can enhance performance in these tasks.

The recurrence of a random walk, defined as the probability that the walker returns to the starting position, was first studied by P\'olya~\cite{Polya1921}. He showed that balanced RWs on an infinite line and plane are recurrent, meaning that the walker returns to the origin with certainty, although the expected return time is infinite. However, for lattices of dimension $d\geq 3$, the walk is transient, and the walker has a non-zero probability of never returning. 

Any investigation of recurrence in quantum settings necessarily includes the definition of the measurement procedure, due to the non-trivial effect of the measurement on the quantum state. Different measurement schemes can result in non-equivalent concepts of recurrence of quantum systems. A scheme was proposed that negates disturbance by measurement for discrete-time~\cite{Stefanak2008} and continuous-time QW~\cite{Darazs2010}, where the measurement is postponed to the end of the evolution and the QW is restarted and iterated for one more step. Another scheme, which is a direct generalization of the P\'olya's concept, was considered by Gr\"unbaum et al~\cite{Grunbaum2013} in a more general setting of iterated unitary evolution. In this case, the return of the walker to the origin is monitored by a partial measurement at each step, and the evolution proceeds only if the walker is not found, yielding a conditional quantum dynamics. Note that when applied to classical RWs the two schemes agree in classifying a walk as recurrent or transient, although in the latter case the recurrence probabilities will generally differ. However, this equivalence is broken in the quantum case, highlighting the role of measurement in quantum mechanics. Both measurement schemes were experimentally implemented in a photonic time-multiplexing setup for a two-state QW on a line~\cite{Nitsche2018} and, more recently, using bulk optics with a single-photon source~\cite{Chen2024}.

We note that, instead of monitoring the return to the origin, one can track a different vertex of the graph to determine whether the quantum walker has reached it. The expected time for the walker to be absorbed at the monitored vertex is referred to as the hitting time. While in some cases the hitting time of a QW can be exponentially faster compared to the classical case \cite{kempe_discrete_2005}, in others QWs can have infinite hitting times \cite{krovi_hitting_2006,krovi_quantum_2006} even for finite graphs, which is not possible for balanced RWs. Both effects are the result of quantum interference of the probability amplitudes, constructive or destructive, respectively.
The concepts of recurrence, return time, and hitting time were also extensively studied in continuous-time QWs \cite{Dhar_quantum_2015,Dhar_detection_2015,Thiel2018,Yin_large_2019,Yin_restart_2023,wang_first_2024}.

In the present paper, we extend the study of recurrence to the discrete-time quantum stochastic walk (DTQSW). DTQSW can be considered as an open quantum system, and we describe it using Kraus operators determined by the underlying quantum and classical dynamics. We focus on recurrence of a particular DTQSW on a line, which is an interpolation between a two-state QW and a balanced RW. We employ the monitored recurrence approach \cite{Grunbaum2013,bourgain_quantum_2014}, which was extended to open system evolution \cite{grunbaum:2018}. By utilizing the first return generating functions \cite{grunbaum:2018}, we approximate the recurrence probability  with significantly higher accuracy than is achievable through finite-time direct numerical simulations. Since the model under consideration interpolates between a transient QW and a recurrent RW, one might expect the recurrence probability of the DTQSW  to exceed that of the unitary QW. Surprisingly, our results show that adding the possibility of taking a classical step can actually decrease the recurrence probability, and we identify the parameter regimes for which this counterintuitive behavior occurs. We also investigate an alternative model of DTQSW, where we interpolate between a two-state QW and a correlated RW, both utilizing the same coin. In this case, we show that if the probability of doing a classical step is nonzero, the DTQSW is recurrent. The crucial difference between the two models is that the latter one involves the measurement of the coin state, causing its decoherence. Our results indicate that any amount of such decoherence enforces the classical behavior resulting in recurrence. 

\section{\label{sec:DTQSW}Discrete-time quantum stochastic walk}

We begin with the notation for the usual unitary two-state discrete-time quantum walk on a line. The Hilbert space of the walk is given by a tensor product of coin and position spaces $\mathcal{H} = \mathcal{H}_c\otimes\mathcal{H}_p$, which are given by
\begin{eqnarray}
    \nonumber \mathcal{H}_p & = & \mathrm{Span}\ \{\ket{x}|\ x\in\mathds{Z}\}, \\
    \mathcal{H}_c & = & \mathrm{Span}\ \{\ket{R},\ \ket{L}\}.
\end{eqnarray}
The time evolution of the discrete-time QW is governed by the equation $\ket{\Psi(t+1)}=\hat{U}\ket{\Psi(t)}$, where $\hat{U}$ is a unitary evolution operator of the form 
\begin{equation}
    \hat{U}~=~\hat{S} \hat{C}.
\end{equation}
The conditional shift operator $\hat{S}$ coherently propagates the quantum walker from its position $x$ according to the coin state
\begin{eqnarray}
\hat S & = & \ketbra*{R}{R}\otimes \hat{T} + \ketbra*{L}{L}\otimes \hat{T}^{\dagger} ,
\label{eq:stepOpr}
\end{eqnarray}
where $\hat{T}=\sum_{x} \ketbra*{x+1}{x}$ is the right displacement operator. For the coin operator we consider the homogeneous case
\begin{equation}
    \hat C = \hat C_\theta \otimes \hat I_p,
\end{equation}
where we parametrize the coin matrix acting on $\mathcal{H}_c$, without loss of generality~\cite{Tregenna2003,goyal_2015_unitary_equiv}, with a single angle $\theta$ according to
\begin{equation}
\hat{C}_{\theta}=
\mqty(\cos\theta &  \sin\theta \\ 
      \sin\theta & -\cos\theta), \quad  \theta\in \left[0,\frac{\pi}{2}\right] \ .
\label{eq:coinOpr}
\end{equation}
For $\theta = \frac{\pi}{4}$ we recover the extensively studied case of the Hadamard walk \cite{ambainis,konno_quantum_2002,grimmett_weak_2004}, which is a quantum analogue of a classical balanced random walk. After $t$ steps the state of the quantum walker takes the general form
\begin{eqnarray}
\ket{\Psi(t)} & = & \hat{U}^t\ket{\Psi(0)} =  \sum_{\substack{
    c=R,L \\
   x \in \mathbb{Z}}}\psi_{c,x}(t)\ket*{c,x},
\end{eqnarray}
where $\psi_{c,x}(t) \in \mathbb{C}$, $c=R,L$ is the amplitude corresponding to the coin state $\ket*{R}$ or $\ket*{L}$, position $x$ and step number $t$. As the initial condition $\ket{\Psi_0}$ we consider a walker localized at the origin
\begin{equation}
    \ket{\Psi(0)} = \ket{\psi_c}\otimes \ket*{0},
\end{equation}
with an arbitrary coin state
\begin{equation}
\begin{gathered}
    \ket{\psi_c} =  a \ket*{R} + b\ket{L}, \\
    a, b \in \mathbb{C}, \quad
    |a|^2 + |b|^2 = 1 .
\end{gathered}
\end{equation}

Let us now turn to the discrete-time quantum stochastic walk (DTQSW) on a line, where we interpolate between the unitary QW and a classical RW. We focus on the case where the classical evolution is a simple balanced RW, where the particle can move to the left or to the right with probability $\frac{1}{2}$. The classical step thus incoherently shifts the whole wave function. The DTQSW model can be understood as an open quantum system that we are going to describe using appropriate Kraus operators. In our case we obtain a completely positive trace preserving (CPTP) map $\cal T$ which acts on the density matrix according to
\begin{align}
{\mathcal T} \hat\rho = \sum_{j=0}^2 \hat E_j \hat\rho \hat E_j^\dagger, 
\label{eq:teDm}
\end{align}
where the Kraus operators $\hat E_j$ read
\begin{eqnarray}
\label{eq:teDm2}    \hat E_0 & = & \sqrt{1-p}\ \hat U , \quad p\in[0,1], \\
\nonumber \hat E_1 & = & \sqrt{\frac{p}{2}}\ \hat I_c\otimes\hat T, \quad \hat E_2 =  \sqrt{\frac{p}{2}}\ \hat I_c\otimes\hat T^{\dagger}     .   
\end{eqnarray}
The operator $\hat E_0$ describes the unitary QW, while $\hat E_1$ and $\hat E_2$ correspond to the classical RW. For $p=0$ the CPTP map $\cal T$ reduces to the unitary QW, while for $p=1$ it becomes a balanced RW. The evolution of the DTQSW from an initial state $\hat \rho(0)$ is then given by
\begin{equation}
    \hat\rho(t) = {\cal T}^{\circ t}\hat\rho(0).
\end{equation}

\section{Recurrence of DTQSW on a line}
\label{sec:rec}
Let us now turn to the central topic of our paper, which is the recurrence of DTQSW on a line. We follow the monitored approach \cite{Grunbaum2013,bourgain_quantum_2014,grunbaum:2018}, where the return of the walk to the origin is detected by the positive outcome of the measurement
\begin{equation}
\hat{\pi}_0 = \hat{I}_c\otimes\ketbra{0}{0}.
\label{eq:sinkAtj}
\end{equation}
The identity operator acting on the coin space implies that we focus on the return to the initial position only, disregarding the coin state, i.e. a recurrence of a subspace \cite{bourgain_quantum_2014}, rather than recurrence of the initial state itself.
The evolution stops if the quantum walker is detected. In the opposite case the evolution continues, with the amplitude at the initial vertex $0$ erased by the complementary projection 
\begin{equation}
\hat \pi_0' = \hat I - \hat\pi_0    . 
\end{equation}
In the density matrix formalism, the projection corresponds to a superoperator $\mathcal{Q}$ which acts as 
\begin{equation}
   \mathcal{Q} \hat \rho = \hat \pi_0' \hat\rho \hat \pi_0' .
   \label{eq:qq}
\end{equation}
A single step of the DTQSW on a line including the measurement at the origin is accordingly described by a completely positive (CP) map $\mathcal{M}$ given by
\begin{eqnarray}
\nonumber \hat\rho(t+1) & = & \mathcal{M} \hat\rho(t) = {\cal Q} \circ {\cal T} \hat\rho(t) \\
\label{eq:one-step}  &=&  \hat{\pi}_0' \qty(\sum_{j=0}^2 \hat E_j \hat\rho(t) \hat E_j^\dagger )\hat{\pi}_0' .
\end{eqnarray}
Due to the projection the trace of $\hat\rho(t)$ is no longer normalized to unity. Instead, it provides the \textit{survival probability} 
\begin{equation}
\label{surv:opens}
S_t = \Tr \hat\rho(t),
\end{equation}
i.e. the probability that the walker was not absorbed until time $t$. The survival probability depends on the coin angle $\theta$ (via $\hat U$) and the probability of doing a classical step $p$. This will be left implicit, for the most part, in the notation. The limiting value of the survival probability $S$ defines the recurrence probability $R$
\begin{equation}
\label{def:rec}
    R = 1 - S = 1 -  \lim\limits_{t\to\infty} S_t .
\end{equation}
For $p=1$ the DTQSW turns into a balanced RW, which was shown to be recurrent by P\'olya \cite{Polya1921}, so $R(p=1) = 1$. In the other limiting case $p=0$, the problem reduces to the recurrence of the unitary QW $\hat U$, which was investigated in the literature \cite{Grunbaum2013,sabri_conditional_2018}. The explicit formula for the two-state QW with the coin (\ref{eq:coinOpr}) is given by
\begin{equation}
\label{surv:p=0}
    R(p=0) =  \frac{2}{\pi}\left[\theta(1- \cot^2{\theta})+\cot{\theta}\right],
\end{equation}
independent of the actual initial coin state of the quantum walker. As we show in Appendices \ref{app:a} and \ref{app:b}, this property is retained for our model of DTQSW for any value of $p$. 

For $p\neq 0,1$ the DTQSW is a nontrivial stochastic quantum evolution and we have to investigate the recurrence of the CPTP map $\cal T$, which is more demanding to evaluate both numerically and analytically. Indeed, for unitary QWs it is sufficient to work with state vectors, whereas for stochastic evolution we have to calculate the density matrix, which increases the required memory allocation quadratically. Moreover, the convergence of the return probabilities is much slower for the DTQSW, complicating the estimation of the final value and even the qualitative assessment of which observed phenomena are just effects of finite simulation time. For these reasons, we utilize the theory of monitored recurrence of open quantum dynamics\cite{grunbaum:2018}, which allows us to evaluate the recurrence probability in an alternative way that does not rely on iterating the CP map (\ref{eq:one-step}).

To study recurrence of the CPTP map $\cal T$ we make use of the first-return generating function (FR-function)
\begin{equation}
\label{f}
    f(z) = (I-\mathcal{Q}) \mathcal{T} (I-z\mathcal{Q}\mathcal{T})^{-1} (I-\mathcal{Q}),
\end{equation}
and the reduced FR-function
\begin{equation}
\label{reduced:f}
\mathcal{F}(z) = \mathcal{P} \mathcal{T} (I - z\mathcal{Q}\mathcal{T})^{-1} \mathcal{P} = \mathcal{P}f(z)\mathcal{P}.
\end{equation}
Here $\mathcal{P}$ denotes the projection onto the origin
\begin{equation}
   \mathcal{P} \hat{\rho} = \hat\pi_0 \hat\rho \hat\pi_0.
\end{equation}
For an initial state $\hat\rho(0)$, the meaning of $\mathcal{F}(z)$ is
\begin{equation}
  \Tr\left[\mathcal{F}(z) \hat\rho(0)\right] = \sum_{n=0}^\infty q_n\left[\hat\rho(0)\right] z^n,
  \label{trace:f:z}
\end{equation}
where $q_n\left[\hat\rho(0)\right]$ denotes the probability of first detection of the walker in its initial position 0 after $n$ steps conditioned on no detections in preceding steps. Therefore, the recurrence probability of the initial subspace is given by
\begin{equation}
\label{rec:prob}
  R = \sum_{n=0}^\infty q_n\left[\hat\rho(0)\right] = \lim\limits_{z\to 1^-} \Tr\left[\mathcal{F}(z)\hat\rho(0)\right].
\end{equation}
The crucial point in the evaluation of the recurrence probability is the renewal equation
\begin{equation}
\label{eq:renew}
f(z) = \frac{1}{z}(I - s(z)^{-1}),
\end{equation}
(the inverse understood to be taken only on the target subspace of the projector $I - \mathcal{Q}$), relating the FR-function $f(z)$ to the Stieltjes function
\begin{equation}
\label{stieltjes}
s(z) = (I-\mathcal{Q}) (I-z\mathcal{T})^{-1} (I-\mathcal{Q}) ,  
\end{equation}
which is much easier to find. Since the QW operator $\hat U$ is homogeneous in position, the inverse appearing in the Stieltjes function can be evaluated using the Fourier transformation, since it acts point-wise in momentum, which is not the case for \eqref{reduced:f}.

\begin{widetext}

Vectorizing the density matrices, the CPTP map $\mathcal{T}$ (\ref{eq:teDm}) corresponds to the operator 
\begin{eqnarray}
\hat{\mathcal{T}} & = & (1-p)\ \hat U \otimes \hat U^* + \frac{p}{2} \left(\hat I_c \otimes \hat T\right)\otimes \left(\hat I_c \otimes \hat T\right)^*  + \frac{p}{2} \left(\hat I_c \otimes \hat T^{-1}\right)\otimes \left(\hat I_c \otimes \hat T^{-1}\right)^*,
\end{eqnarray}
where the asterisk denotes complex conjugation of the operator (in the working basis). As all involved operators have all matrix elements real-valued, a simpler form holds,
\begin{eqnarray}
\hat{\mathcal{T}} & = & (1-p)\ \hat U \otimes \hat U + \frac{p}{2} \left(\hat I_c \otimes \hat T\right)\otimes \left(\hat I_c \otimes \hat T\right)  + \frac{p}{2} \left(\hat I_c \otimes \hat T^{-1}\right)\otimes \left(\hat I_c \otimes \hat T^{-1}\right)
\label{eq:t-vec}
\end{eqnarray}
In the momentum representation $\hat{\mathcal{T}}$ takes the form
\begin{equation}
\hat{\mathcal{T}}  =  \int\limits_0^{2\pi} \frac{dk_1}{2\pi} \int\limits_0^{2\pi}  \frac{dk_2}{2\pi}\ V(k_1,k_2)\otimes \ketbra{k_1,k_2}{k_1,k_2},
\end{equation}
where we have reshuffled the ordering of the Hilbert spaces to simplify the formula. By $V(k_1,k_2)$ we have denoted the operator acting on the coin spaces
\begin{eqnarray}
   V(k_1,k_2) & = &   (1-p) U(k_1)\otimes U(k_2) +  p\cos(k_1+k_2) I_c\otimes I_c,      
\end{eqnarray}
where $U(k)$ describes the quantum step in the Fourier picture
\begin{align}
U(k) &= \mathrm{diag}(e^{-i k},\ e^{ik}) \cdot C_\theta .
\end{align}
The resolvent can be expressed in the form
\begin{equation}
(I - z\hat{\mathcal{T}})^{-1} = \int\limits_0^{2\pi} \frac{dk_1}{2\pi} \int\limits_0^{2\pi} \frac{dk_2}{2\pi} A(z,k_1,k_2) \otimes \ketbra{k_1,k_2}{k_1,k_2},
\end{equation}
where we introduce the notation
\begin{eqnarray}
\label{Az:k} A(z,k_1,k_2)  & = & \left[I_c\otimes I_c - z V(k_1,k_2)\right]^{-1} = \left[(1-z p\cos(k_1+k_2)) I_c\otimes I_c -  z(1-p) U(k_1)\otimes U(k_2)\right]^{-1}.
\end{eqnarray}
Incorporating the projections
\begin{equation}
\hat I - \hat{\mathcal{Q}} = \hat I_c\otimes \hat I_c\otimes\left( \hat I_p\otimes\ketbra{0}{0} + \ketbra{0}{0}\otimes  \hat I_p - \ketbra{0,0}{0,0}\right),
\end{equation}
we obtain the Stieltjes function as a sum 
\begin{align}
    \label{eq:stieltjes}
    \hat s(z)  = & \sum\limits_{\substack{
    x,y,m,n \\
    xm=yn=0}} A_{xm,yn}(z)\otimes \ketbra{x,m}{y,n}
\end{align}
of operators defined by
\begin{equation}
A_{xm,yn}(z) = \int\limits_0^{2\pi} \frac{dk_1}{2\pi} \int\limits_0^{2\pi} \frac{dk_2}{2\pi} A(z,k_1,k_2) e^{i k_1 (x-y)+i k_2 (m-n)}
\label{eq:rxy-fourier}
\end{equation}
\end{widetext}
The integrals have to be evaluated numerically, for more details we refer to Appendix \ref{app:c}. From the form of \eqref{eq:t-vec} together with the bipartiteness of the walking graph (also honored by the incoherent shifts) it follows that $A_{xm,yn}(z)$ is zero when $x+y+m+n$ is odd, partitioning the operator into two invariant subspaces within $\mathop{\textrm{Im}} (I - \mathcal{Q})$. The vectorized initial state is fully contained in one of them,
\begin{equation}
    \mathop{\mathrm{Span}}\left\{ \ket{c,c',x,m} \middle| \begin{aligned}
    &c,c' \in \{\ket{R}, \ket{L}\}, \\ &x,m \text{ even}, x = 0 \vee m = 0
    \end{aligned}\right\},
    \label{eq:subspace}
\end{equation}
so it suffices to take the inverse on this invariant subspace only for the purposes of calculating $R$ using \eqref{rec:prob}.

Since (\ref{eq:rxy-fourier}) has to be treated numerically, we do not have analytical formulas for the Stieltjes function and the FR-functions in the parameter $z$. However, we can evaluate ${\cal F}(z)$ at some $z \lesssim 1$ and thus obtain an approximation of the recurrence probability as given by \eqref{trace:f:z},
\begin{equation}
  \tilde R_z := \Tr\left[\mathcal{F}(z) \hat\rho(0)\right] = \sum_{n=0}^\infty q_n\left[\hat\rho(0)\right] z^n.
  \label{rec:prob:z}
\end{equation}
This can be put in contrast with the direct evaluation of the evolution (\ref{eq:one-step}) using the CP map $\cal M$, from which the return probability after $t$ steps
\begin{equation}
\label{rec:prob:t}
  R_t = 1 - S_t = \sum_{n=0}^t q_n\left[\hat\rho(0)\right] = \sum_{n=0}^\infty q_n\left[\hat\rho(0)\right] H(t-n),
\end{equation}
can be found. Here, $H(x)$ is the Heaviside theta function with $H(0) = 1$. Apparently, the two approaches differ in the weight factors assigned to the first-return probabilities $q_n$, both approaching constant unity for \eqref{rec:prob} from below, but in different ways. In order to put them in comparison, we may, however, consider the characteristic time $t$ at which $z^t$ decreases to $1/e$, or we take a number of terms in \eqref{rec:prob:t} such that the total of the weight factors agrees. For $z \lesssim 1$, both estimates coincide at an \emph{effective} number of $t_{\mathrm{eff}} = 1/(1-z)$ steps. As we discuss in the Appendix \ref{app:c}, the numerical evaluation of $\tilde R_z$ remains stable up to around $z=0.99999$, corresponding to $t_{\mathrm{eff}} = 10^5$, surpassing the number of iterations of the CP map $\cal M$ reached by direct evaluation by several orders of magnitude.

\section{Results}
\label{sec:discussion}

\begin{figure}[t]
\includegraphics[width=\linewidth]{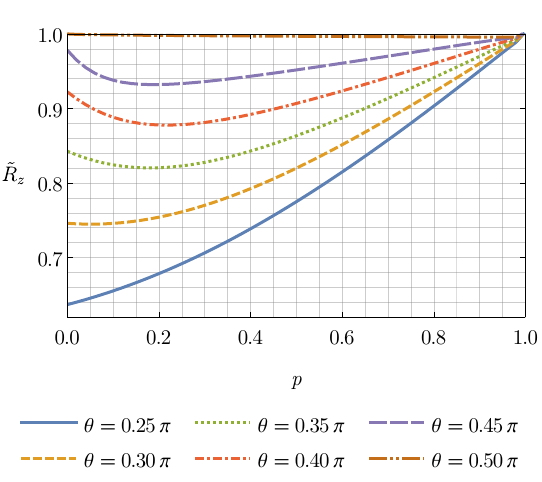}
\caption{Estimate of recurrence probability evaluated with the generating function (\ref{rec:prob:z}) for different values of the coin angle $\theta$ as a function of the parameter $p$. We used $z=0.99999$, comparable to $t_{\mathrm{eff}} = 10^5$ steps. }
\label{fig:results}
\end{figure}

We present the results for recurrence probability of DTQSW evaluated using the generating function (\ref{rec:prob:z}) in Figure~\ref{fig:results}, where we plot the recurrence probability approximated by $\tilde{R}_{0.99999}$ as a function of the interpolation probability $p$ for different values of the coin angle $\theta$.
Our convergence analysis, presented in full in Appendix \ref{app:c}, shows that this approximant should not differ from the actual value of $R$ by more than $10^{-2}$ in any data point $(\theta, p)$.
The results show that for smaller values of $\theta$, such as $\theta=\frac{\pi}{4}$ corresponding to the Hadamard walk (blue curve), the recurrence probability is a monotonously increasing function of $p$. However, for $\theta > \theta_* \approx 0.289 \pi$ we observe that $\tilde{R}_z(p)$ becomes non-monotonous as it decreases for small values of $p$. This is a rather counterintuitive behavior, since the DTQSW interpolates between a transient QW (i.e. $R(p=0)<1$ for $\theta\neq \frac{\pi}{2}$) and a recurrent balanced RW (i.e. $R(p=1)=1$ independent of $\theta$).

\begin{figure}[h!]
{\leavevmode
\vtop{\vskip0pt\hbox to 0pt{(a)\hss}}%
\includegraphics[width=0.47\textwidth]{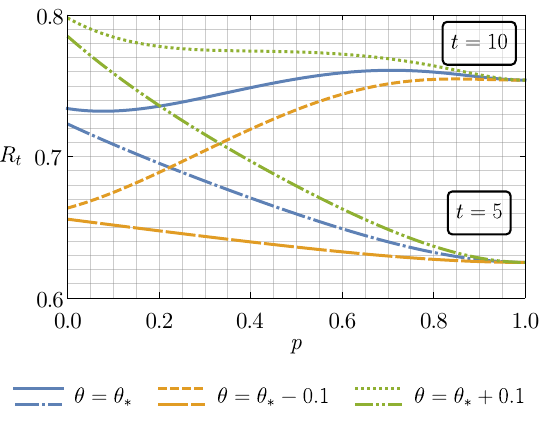}\hfill
\vtop{\vskip0pt\hbox to 0pt{(b)\hss}}%
\includegraphics[width=0.47\textwidth]{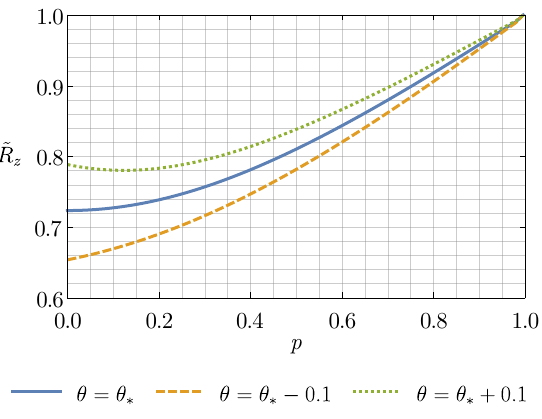}}
\caption{Recurrence probability for $\theta_*$ (blue), $\theta_* - 0.1\,\text{rad}$ (orange) and $\theta_* + 0.1\,\text{rad}$ (green), where the estimate $0.2892\pi$ was substituted for the unknown exact value. In (a) we show the return probability after 5 steps (lower lines in the legend) and 10 steps (higher three lines). For 5 steps all three curves are decreasing as functions of $p$. At 10 steps, all return probability values increase, with the change more pronounced with higher $p$, and likewise do the derivatives at $p = 0$ of all three displayed cases. The 10-step return probability becomes an increasing function for $\theta=\theta_*-0.1$, while the other two cases maintain a negative slope near $p = 0$. (b) shows the final recurrence probability estimated using the generating functions with $z=0.99999$. One can see that for $\theta>\theta_*$ the initial decrease survives while the rightmost value approaches 1, forming a dip.}
\label{fig:rec:critical}
\end{figure}

As we illustrate in Figure~\ref{fig:rec:critical}, the dip in the recurrence probability develops in the first few steps. In Figure~\ref{fig:rec:critical}(a), we display $R_t(p)$ for $t=5$ and $t=10$ steps for three choices of the coin angle given by $\theta_*$, $\theta_* -0.1\,\text{rad}$ and $\theta_* + 0.1\,\text{rad}$. For $t=5$, all three curves decrease with $p$. Indeed, $R_5(p=1)<R_5(p=0)$ for these values of $\theta$, i.e. doing a classical step decreases the return probability compared to the unitary QW in the early steps. However, as the number of steps increases, $R_t(p=1)$ tends to 1. Already for 10 steps $R_{10}(p)$ becomes an increasing function of $p$ for $\theta = \theta_* - 0.1$. Nevertheless, for $\theta = \theta_* + 0.1$, $R_{10}(p)$ maintains the dip for small $p$. Figure~\ref{fig:rec:critical}(b), where we display the recurrence probability estimated using the generating function (\ref{rec:prob:z}), shows that the dip persists in the asymptotic regime. Indeed, due to the quantum interference the DTQSW spreads ballistically for $p\neq 1$ and hence it is transient. The counterintuitive behavior of the recurrence probability for $\theta>\theta_*$ thus truly emerges from the interplay of the classical balanced RW and the QW.

\begin{figure}
\includegraphics[width=\linewidth]{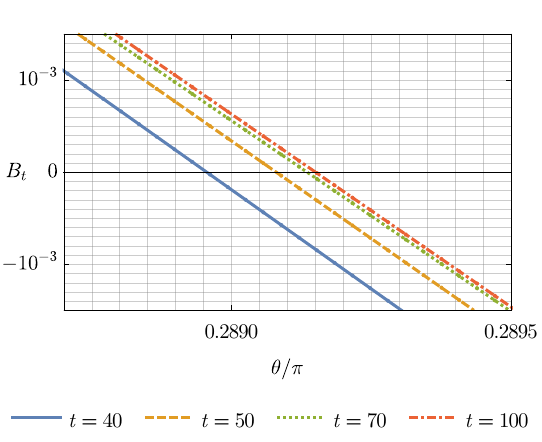}
\caption{Numerically computed first derivative of $R_t$ at zero $R_t'(p=0) = B_t$ against the coin angle $\theta$ with focus on the parameter range where $B_t$ crosses zero, with $t$ chosen to grow in increasing steps. The simulation indicates that $\theta_* \gtrsim 0.28915 \pi$.}
\label{fig:B100}
\end{figure}

Concerning the cross-over point $\theta_*$, where the non-monotonicity of $R(p)$ emerges, the evaluation of the generating function provides a value $\theta_* \approx 0.2892 \pi$. We also determine it following an alternative approach based on first order perturbation expansion of the CP map ${\cal M}$ (\ref{eq:one-step}). This allows us to evaluate the first derivative of the return probability at $p=0$ $R_t'(p=0) = B_t$ efficiently; for details we refer to the Appendix \ref{app:d}. A close examination of the cross-over point is presented in Figure~\ref{fig:B100}, where we consider different values of $t$ up to $100$. The numerical evaluation of the first derivative $B_t$ gives the transition at $\theta_* \approx 0.28915\pi$. While yet higher number of steps may push the value somewhat higher still, this result is in good agreement with the result from the evaluation of the generating function.

The behavior depicted in Figure~\ref{fig:results} can be put in contrast with a different DTQSW model, where we interpolate between a QW and the classically correlated RW, which also uses a coin state (although classical) to keep track of the direction taken in the preceding step. The resulting CPTP map contains five Kraus operators, one for the QW and four for the correlated RW. The details are described in Appendix \ref{app:f}. In Figure~\ref{fig:recur:crw} we present its return probability $R_t(p)$ for the coin angle $\theta = \frac{2\pi}{5}$ as a function of the classical step probability $p$. The results were obtained by directly simulating the monitored evolution of the corresponding CPTP map. We see that $R_5(p)$ (blue line) is independent of $p$, indicating that quantum interference does not affect the return probability for the first 5 steps. For $t=10$ (orange dashed curve), the return probability is a monotonously increasing function of $p$. With an increasing number of steps (the green dotted curve corresponds to $t=100$), $R_t(p)$ tends closer to 1 for $p\neq 0$ and the slope at $p=0$ is steeper. The same behavior is obtained for all values of $\theta\neq 0$, i.e. $R_t(p)$ is always an increasing function of $p$ for $t>5$. 

\begin{figure}
\includegraphics[width=\linewidth]{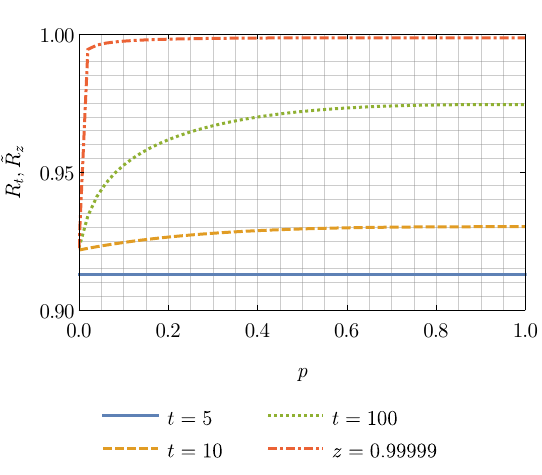}
\caption{Recurrence probability of the DTQSW where we interpolate between a QW and the correlated random which has the same coin angle $\theta$. We have chosen $\theta = \frac{2\pi}{5}$. Blue line is obtained for $t=5$ steps, orange dashed curve corresponds to the $t=10$ steps and green dotted curve to $t=100$ steps. For the first 5 steps the return probability is independent of $p$. After 5 steps the dependence on $p$ emerges, however, $R_t(p)$ is always an increasing function. Notice that with increasing number of steps $t$ the slope at $p=0$ becomes steeper. The red dot-dashed curve is evaluated using the generating functions. The data correspond to the choice $z=0.99999$, i.e. $t_\text{eff} = 10^5$. Asymptotically, the leftmost point is pinned by \eqref{surv:p=0}, which for the used $\theta$ predicts $R \approx 0.9224$, but all points at $p > 0$ tend to 1.}
\label{fig:recur:crw}
\end{figure}

Evaluating the recurrence probability using the generating functions approach, we find that $R(p)$ approaches 1 for $p\neq 0$, see the red dot-dashed curve in Figure~\ref{fig:recur:crw} which depicts the $\tilde{R}_z(p)$ for $z=0.99999$. This indicates that the recurrence probability is in fact a step function with $R(p) = 1$ for $p>0$ $R(p=0)<1$ being given by the unitary case (\ref{surv:p=0}). The difference between the two models of DTQSW is in the Kraus operators describing the classical walks. For the balanced RW, $\hat E_{1,2}$ given in equation (\ref{eq:teDm2}) act independently of the coin. However, the Kraus operators for the correlated RW involve the measurement of the coin state. These projections cause decoherence of the coin state, which fundamentally affects the recurrence of the corresponding DTQSW. Indeed, our results indicate that any amount of such decoherence will ultimately result in classical behavior, i.e. the recurrence probability will be one for $p\neq 0$, $\theta \neq 0$.

\section{Discussion}
\label{sec:conclusions}

The recurrence of DTQSW on a line has been investigated in detail. We have shown that adding the possibility of doing a balanced RW can decrease the recurrence probability, despite the fact that the classical evolution is recurrent. Suppression of recurrence probability occurs for DTQSWs with coin parameter within the interval $\theta\in (\theta_*,\frac{\pi}{2})$, where the threshold value is numerically identified as $\theta_* \approx 0.2892 \pi$.Utilizing the first return generating functions we conclude that the observed results are not due to finite-size effects but are features of the asymptotic evolution. In line with earlier findings, which showed that certain levels of stochasticity in continuous-time quantum walks may provide algorithmic advantage, our results indicate that classical randomness may be beneficial in tasks using discrete-time quantum walks and partial measurement, such as subroutines of quantum algorithms.

Recurrence of DTQSW crucially depends on the particular choice of the quantum stochastic evolution. The model interpolating between QW and correlated RW contains Kraus operators which include measurement of the coin state, and the corresponding DTQSW is recurrent for any non-zero amount of stochasticity. Such measurements lead to decoherence of the coin state, which in turn enforces the classical behavior of recurrence. We aim to investigate if this holds for different models of decoherence as well, and obtain a better understanding between coherence and recurrence probability of DTQSW. In addition, more general models of DTQSW can be considered, which do not have the form of interpolation between quantum and classical evolution. An interesting example can be an extension of classical pursuit problems \cite{krapivsky_kinetics_1996,redner_capture_1999} to the quantum domain. One can also consider DTQSW with different QW evolution with higher-dimensional coin space, e.g. lazy QW on a line or QW on higher-dimensional lattices. In such a case, the unitary QW already shows richer behavior, since the recurrence probability can depend on the initial coin state \cite{bourgain_quantum_2014,stefanak_rec_2023}. Finally, it is possible to focus on the recurrence of the exact initial state \cite{Grunbaum2013}, i.e., including the initial state of the coin. For some unitary QWs this can lead to an apparent paradox where the state recurrence is greater than the site recurrence \cite{bourgain_quantum_2014,stefanak_rec_2023}. We aim to extend these investigations to quantum stochastic walks in the near future.

Finally, let us briefly comment on the possible experimental realizations of the observed phenomena. Investigation of recurrence requires to perform partial projective measurements without disturbing the rest of the superposition state. Such measurements were demonstrated in photonic implementations of QWs \cite{Nitsche2018,Chen2024}, and in quantum processors based on superconducting qubits \cite{wang_first_2024} utilizing the mid-circuit readout \cite{corcoles_exploiting_2021}. Stochastic evolution in quantum processors can be introduced by employing ancillary qubits, whose values control unitaries performed on the quantum walk register qubits. Ancillary qubits have to be appropriately rotated and measured before every step. However, already implementations of unitary discrete-time QWs are highly non-trivial due to interaction between coin and position register, resulting in large depth already for a single step unitary evolution operator \cite{acasiete_implementation_2020,singh_quantum_2021,razzoli_efficient_2024,sarkar_quantum_2024}. Incorporation of sink and stochastic evolution based on mid-circuit measurements, classical feedforward and controlled operations, increases both the width and the depth of the circuit further. Given the noise present in particular in two-qubit operations on current NISQ chips, it is challenging to reach sufficient number of steps to demonstrate the suppression of recurrence probability before the decoherence hinders quantum superpositions. On the other hand, dedicated photonic implementations, in particular the time-multiplexing feedback loop \cite{Nitsche2018}, have long coherence times and allow to reach several tens of steps including partial measurements. Stochasticity can be introduced by utilizing fast-switching programmable devices such as EOMs, which implement coin and deterministic out-coupling of the light pulse into the detection unit. Repeating the experiment with different settings of EOMs one can perform a Monte-Carlo simulation of the chosen DTQSW dynamics. Hence, while challenging, the proposed phenomena are within experimental reach. 

\section*{Acknowledgments}

\.{I}. Y., V. P., M. \v{S}. and I. J. have been supported by the Grant Agency of the Czech Republic GAČR under Grant No. 23-07169S. V.P. and \.{I}.Y. were also supported by the Grant Agency of the Czech Republic under Grant No. 19-15744Y. The support of the European Union's research and innovation program under EPIQUE Project GA No. 101135288 is acknowledged. The funders played no role in study design, analysis and interpretation of data, or the writing of this manuscript. 

\bibliography{bibliography}
\bibliographystyle{quantum}

\appendix

\section{Freedom of restricting the coin to a real-valued matrix}
\label{app:a}

Let us prove that the choice of the real-valued matrix \eqref{eq:coinOpr} is without loss of generality. This freedom is routinely exercised in unitary quantum walks\cite{Tregenna2003,goyal_2015_unitary_equiv}, but it remains to show that it does not affect the result even in our considered case of stochastic evolution and partial measurements.

A general $\mathsf{U}(2)$ matrix may be written as
\begin{equation}
    \hat{\tilde C} = e^{i \phi} e^{i \alpha \hat\sigma_z} \hat C_\theta e^{i \beta \hat\sigma_z},
\end{equation}
which is just a slight modification of the Pauli decomposition. Let us consider such matrix being used instead of $\hat C_\theta$ in the construction of the superoperator $\mathcal{M}$ of \eqref{eq:one-step},
\begin{equation}
    \begin{aligned}
        \tilde{\mathcal{M}}: \hat\rho &\mapsto 
        \hat{\pi}_0' \hat{\tilde E}_0 \hat\rho(t) \hat{\tilde E}_0^\dagger \hat{\pi}_0' \\
        &\quad + \hat{\pi}_0' \hat E_1 \hat\rho(t) \hat E_1^\dagger \hat{\pi}_0' \\
        &\quad + \hat{\pi}_0' \hat E_2 \hat\rho(t) \hat E_2^\dagger \hat{\pi}_0' \\
    \end{aligned}
    \label{eq:m-tilde}
\end{equation}
where
\begin{equation}
\begin{aligned}
    \hat{\tilde E}_0 &= \sqrt{1-p} \hat S (\hat{\tilde C} \otimes \hat I_p) \\
    &= \sqrt{1-p} \hat S \left((e^{i \phi} e^{i \alpha \hat\sigma_z} \hat C_\theta e^{i \beta \hat\sigma_z}) \otimes \hat I_p\right).
\end{aligned}
\end{equation}
Consider the position observable
\begin{equation}
    \hat X = \sum_x x \dyad{x}
\end{equation}
and the operator
\begin{equation}
    e^{i \omega \hat X} = \sum_x e^{i \omega x} \dyad{x}.
\end{equation}
It follows from the definition \eqref{eq:stepOpr} that
\begin{equation}
    (\hat I_c \otimes e^{i \omega \hat X}) \hat S = \hat S (e^{i \omega \hat\sigma_z} \otimes e^{i \omega \hat X}).
\end{equation}
Take now $\omega = \alpha + \beta$ and consider the unitary similarity transform
\begin{equation}
\begin{aligned}
    &(\hat I_c \otimes e^{-i (\alpha + \beta) \hat X}) \hat{\tilde E}_0 (\hat I_c \otimes e^{i (\alpha + \beta) \hat X}) ={} \\
    &= e^{i\phi} \sqrt{1-p} \hat S \left( e^{-i (\alpha + \beta) \hat \sigma_z} e^{i \alpha \hat\sigma_z} \hat C_\theta e^{i \beta \hat\sigma_z}) \otimes \hat I_p \right) \\
    &= e^{i\phi} (e^{-i \beta \hat\sigma_z} \otimes \hat I_p) \hat E_0 (e^{i \beta \hat\sigma_z} \otimes \hat I_p)
\end{aligned}
\end{equation}
Hence
\begin{equation}
    \begin{aligned}
        \hat{\tilde E}_0 &= e^{i\phi} \hat W \hat E_0 \hat W^\dagger, \\
        \hat W &= e^{-i \beta \hat\sigma_z} \otimes e^{i (\alpha + \beta) \hat X}.
    \end{aligned}
\end{equation}
Similarly, we have
\begin{equation}
    \begin{aligned}
        \hat W \hat E_1 \hat W^\dagger &= e^{i(\alpha + \beta)} \hat E_1, \\
        \hat W \hat E_2 \hat W^\dagger &= e^{-i(\alpha + \beta)} \hat E_2.
    \end{aligned}
\end{equation}
Noting that $\hat \pi'_0$ commutes with $\hat W$, and that any global phase factor cancel out in \eqref{eq:m-tilde}, we have
\begin{equation}
\begin{aligned}
    \tilde{\mathcal{M}}(\hat \rho) &= \hat W \mathcal{M} (\hat W^\dagger \hat\rho \hat W) \hat W^\dagger, \\
    \tilde{\mathcal{M}}^n(\hat \rho) &= \hat W \mathcal{M}^n (\hat W^\dagger \hat\rho \hat W) \hat W^\dagger.
\end{aligned}
\end{equation}
The final important observation is that for initial states localized at $x = 0$, the similarity transform $\hat W$ induces a rotation on the coin state only. Thus, as far as the survival probability \eqref{surv:opens} is concerned, the parameters $\phi$ and $\alpha$ have no observable effect, and $\beta$ can be transferred to the initial coin state, after which the phase-less form $\hat C_\theta$ can be used instead of $\hat{\tilde C}$. In the next Appendix, we prove that the change in the initial coin state does not reflect in the result either.

\section{Independence of the recurrence probability on the initial coin state}
\label{app:b}

Here we prove that the recurrence probability of the DTQSW (\ref{eq:teDm}) is independent of the initial coin state. To see this, we rewrite the survival probability (\ref{surv:opens}) in the form
\begin{equation}
\begin{aligned}
    S_t &= \Tr \left( \mathcal{M}^{\circ t}[\hat \rho(0)] \right)
    = \Tr \left( \hat\rho(0) (\mathcal{M}^*)^{\circ t}[\hat I] \right) \\
    &= \Tr \left( \hat\rho_c \prescript{}{p}{\langle 0|(\mathcal{M}^*)^{\circ t}[\hat I]|0\rangle}_{\mathstrut p} \right)
\end{aligned}
\end{equation}
where $\mathcal{M}^*$ is the dual of the superoperator $\mathcal{M}$, mapping
\begin{equation}
    \hat A \mapsto \mathcal{M}^*[\hat A] = \sum_{j=0}^2 \hat E_j^\dagger \hat \pi'_0 \hat A \hat \pi'_0 \hat E_j,
\end{equation}
and $\hat \rho(0) = \hat\rho_c \otimes \ket{0}_p\!\bra{0}$. Next, we will utilize symmetries of QWs familiar e.g. from studies of topological phases \cite{kitagawa_observation_2012,asboth_symmetries_2012,cedzich_bulk-edge_2016,asboth_short_2016,barkhofen_measuring_2017}.
QW with the coin \eqref{eq:coinOpr} has parity symmetry given by the operator
\begin{equation}
    \hat R: \ket{x, c} \mapsto \ket{-x} \otimes \hat \sigma_y \ket{c}
\end{equation}
and symmetry with respect to complex conjugation in the established basis, which we can denote by an anti-unitary $\hat \Upsilon$.
These two symmetries are inherited by the superoperator $\mathcal{M}$, as
\begin{equation}
\begin{aligned}\relax
    [\hat R, \hat E_0] &= 0, \\
    \hat E_{1,2} \hat R &= \hat R \hat E_{2,1}, \\
    [\hat \Upsilon, \hat E_j] &= 0\ \forall j, \\
    [\hat R, \hat \pi'_0] &= [\hat \Upsilon, \hat \pi'_0] = 0.
\end{aligned}
\end{equation}
The identity operator $\hat I$ also has both these symmetries, so it must also be true for $(\mathcal{M}^*)^{\circ t}[\hat I]$.
Now $\prescript{}{p}{\langle 0|(\mathcal{M}^*)^{\circ t}[\hat I]|0\rangle}_{\mathstrut p}$ is an operator $\hat T$ acting on $\mathcal{H}_c$ only whose matrix needs to commute with $\hat\sigma_y$, be real-valued and positive (hence self-adjoint). The first condition restricts the possibilities to linear combinations of $\hat\sigma_y$ and the identity matrix,
\begin{eqnarray}
    \hat T = \mu I + \nu \hat\sigma_y.
\end{eqnarray}
The second condition constrains $\mu$ to real values and $\nu$ to pure imaginary, while self-adjointness would require $\nu$ to be real, too. Together with positivity this only leaves \begin{equation}
\begin{aligned}
    \hat T &= \mu \hat I_c, \quad \mu \ge 0, \\
    S_t = \Tr(\hat\rho_c \hat T) &= \mu \Tr(\hat\rho_c) = \mu,
\end{aligned}
\end{equation}
i.e. the survival probability is independent of the initial coin state, and so is the return probability (\ref{def:rec}).

\section{Evaluation of the generating functions}
\label{app:c}

We discuss in detail the numerical evaluation of the Stieltjes function (\ref{stieltjes}) that is used to determine the FR-function (\ref{eq:renew}). The formula (\ref{eq:stieltjes}) expresses $\hat s(z)$ as a direct sum of operators (\ref{eq:rxy-fourier}), which are given by the Fourier transform of matrices $A(z,k_1,k_2)$ defined by (\ref{Az:k}). We evaluate the inverse in (\ref{Az:k}) with the help of the adjugate matrix
\begin{equation}
\label{app:amat}
    A(z,k_1,k_2) = \frac{1}{D} \operatorname{adj}(I_c\otimes I_c - z V(k_1,k_2)),
\end{equation}
where $D$ is the determinant 
\begin{eqnarray}
  D & = &  \det(I_c\otimes I_c - z V(k_1,k_2)) \\
  \nonumber & = & (\rho^2-\sigma^2)^2 + \left[2 \rho \sigma (1-\cos{\xi} \cos{\eta}) - \right.\\
\nonumber   & & \left. - \left(\rho^2+\sigma^2\right) (\cos{\xi}-\cos{\eta})\right] 2 \rho \sigma \cos^2\theta. 
\end{eqnarray}
Here we have used the notation
\begin{eqnarray}
 \xi & = & k_1 + k_2, \quad \eta = k_1 - k_2 ,\\
 \nonumber \rho & = & 1 - z p\cos\xi , \quad \sigma = z(1-p) .
\end{eqnarray}
Note that the matrix $I_c\otimes I_c - z V(k_1,k_2)$ consists of linear combinations of functions $e^{\pm ik_1} e^{\pm ik_2}$. By construction, the matrix elements of the adjugate matrix are Laurent polynomials in $e^{ik_1}$, $e^{ik_2}$. Hence, the decomposition (\ref{app:amat}) allows us to calculate the Fourier transform as indicated by \eqref{eq:rxy-fourier} once, for the $1/D$ factor, and obtain all 16 matrix elements using discrete convolution.

The superoperator $\hat s(z)$ of \eqref{eq:stieltjes} acts on an infinite-dimensional space $\Im(\hat I - \hat{\cal Q})$, on which its inverse would need to be taken. We reduce the problem by constraining to \eqref{eq:subspace}, which, nevertheless, is still of infinite dimension. To make this possible numerically, we clamp the position indices $x$, $x'$ within certain $-N_\text{max}$ and $N_\text{max}$. This is equivalent to replacing the projector $I - \mathcal{Q}$ in \eqref{stieltjes} by $\mathcal{B} (I - \mathcal{Q})$ where $\mathcal{B}: \hat\rho \mapsto \hat B \hat\rho \hat B$, $\hat B$ being a projector on the corresponding part of the state space. When such amended operator is used in the renewal equation \eqref{eq:renew}, all occurrences of the projector $\mathcal{Q}$ appearing in \eqref{f} and \eqref{reduced:f} change accordingly to
\begin{equation}
    \tilde{\mathcal{Q}} = I - \mathcal{B} (I - \mathcal{Q}).
\end{equation}
Apart from reducing the result to the same working subspace, this has the effect of describing a slightly modified dynamics. As $\tilde{\mathcal{Q}}$ appears instead of $\mathcal{Q}$ in the central resolvent of \eqref{f}, the values of such modified FR function describe an evolution in which instead of zeroing all elements of $\hat\rho$ corresponding to position 0 in either the bra- or the ket-part in one step of \eqref{eq:one-step}, off-diagonal blocks for which the other index of the matrix exceeds $N_\text{max}$ in absolute value are left intact. Surely this is an unphysical operation as it does not preserve positivity. However, we assume that coherences of the form
\begin{equation}
    \prescript{}{p}{\bra{0}\hat\rho\ket{x}}_{\mathstrut p}, \quad
    \prescript{}{p}{\bra{x}\hat\rho\ket{0}}_{\mathstrut p}, \quad
    |x| > N_\text{max}
\end{equation}
are either ignorable or shielded by other means from interfering with the properties studied here for $N_\text{max}$ sufficiently large.

To optimize speed, the numerical calculation of the Fourier transform was done on a GPU. Further measures were taken to mitigate some pathological properties of the $1/D$ expression for $z$ very close to $1$, namely large areas close to zero lined by sudden sharp crests near where $k_1 \pm k_2$ are integer multiples of $\pi$. These measures included changing the integration coordinates from $(k_1, k_2)$ to $(\xi, \eta)$, thus aligning the irregularities with the coordinate axes, and further substituting the integration variable $\xi$ by $\xi = x - \frac12 \sin(2x)$ and likewise for $\eta$, taking denser samples near the peaks as well as reducing their height as a result of multiplication by $\mathrm{d}\xi / \mathrm{d}x = 1 - \cos(2x)$. After this regularization, the integration was divided into 1024 equal subintervals on each axis and summed by the rectangle rule. Cursory tests showed that higher subdivisions or more advanced integration rules did not improve the results appreciably.

The matrices $A_{xm,yn}(z)$ were calculated for each $\theta,p$ for only those $x, y, m, n$ where $xm = yn = 0$, all involved indices are even and limited between $-N_\text{max}$ and $N_\text{max}$, where a value of 20 was used for $N_\text{max}$ and 0.99999 for $z$ (comparable to considering $10^5$ steps of the walk) based on a convergence analysis presented below. After this step, the trace appearing in \eqref{rec:prob} could be calculated as a sum of just two elements of the inverse of the considered submatrix (as corresponding to $\rho(0) = \ket{R}\bra{R} \otimes  \ket{0}\bra{0}$, leveraging the earlier observation that the recurrence probability does not depend on the initial coin state). Since at this point the matrices were still rather large, $84\times84$, these two individual inverse matrix elements were calculated using Cramer's rule. Both the determinant and the corresponding subdeterminants were simultaneously found by appropriately suited Gaussian elimination, reducing the matrix to $2\times2$ while maintaining maximum numerical stability by choosing the largest possible pivot in each step.

\begin{figure}
    \centering
    \includegraphics[width=\linewidth]{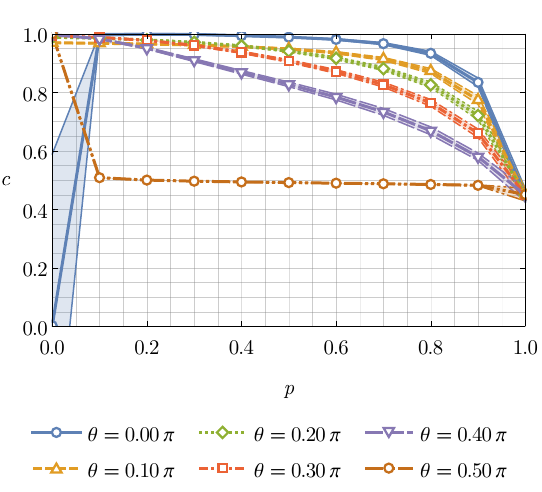}
    \caption{Fit results for varying $z$ parameter. The fitting function is $a - b(1-z)^c$. The best fit value and standard error of the exponent $c$ are plotted for various combinations of $\theta$ and $p$. The following values of $z$ were sampled: 0.99, 0.995, 0.998, 0.999, 0.9995, 0.9998, 0.9999, 0.99995, 0.99998, 0.99999, and $N_{\text{max}}$ was set to 20.}
    \label{fig:convergence-z-in-p}
\end{figure}

Using all these optimizations, the calculation time was reduced to below one second per $(\theta, p, z, N_\text{max})$ data point, running on a consumer-grade integrated GPU. This allowed us to assess the justification of the finite approximations to the unreachable limits $z \to 1_-$, $N_\text{max} \to \infty$. Our analysis as shown in Fig.~\ref{fig:convergence-z-in-p} shows that the speed of convergence in $1-z$ fits well on $a - b (1-z)^c$ with $c$ between 1, characteristic of a QW, and $0.5$, as expected of the return probabilities of a classical RW.

The fit failed in $\theta = 0$, $p = 0$, where the numerical data were displaying some unpredictable noise around zero, but the return probability can easily be shown to be exactly zero, as the walker only goes in a single direction. In all other cases, the limit value can be extrapolated with high confidence. The convergence is slowest for $p = 1$ and for $\theta = \pi/2$, $p > 0$, which are both unbiased classical random walk scenarios with $R = 1$. The former case is well known, the calculation for the latter is presented in the following subsection. ($\theta = 0$, $p > 0$ is also classical but biased, and $\theta = \pi/2$, $p = 0$ is deterministic.) However, with the fitted $b$ parameter being always less than 3 in magnitude, the value at $z = 0.99999$ is representative of the limit to 2 significant digits in all cases, and up to 4 significant digits where the exponent $c$ approaches one. (In choices of $z$ even closer to 1, the limitations of the double-precision floating-point arithmetic start visibly manifesting.)

In terms of $N_\text{max}$, the convergence is more complicated but the differences between the results obtained for $N_\text{max}$ between 20 and 80 in steps of 10 never exceed $2\cdot 10^{-5}$. For this reason $N_\text{max} = 20$ has been chosen as a good representative of the final value within the accepted accuracy, even though rising $N_\text{max}$ did not bring any complications beyond longer calculation times (as opposed to $z \to 1_-$).

The source code used for the numerical evaluation of the generating functions as well as the raw data used to plot the figures presented in the paper are available at the Gitlab repository \cite{dtqsw_code}.

\section{Initial slope of the recurrence probability}
\label{app:d}

\begin{figure}[t]
\centering
\includegraphics[width=\linewidth]{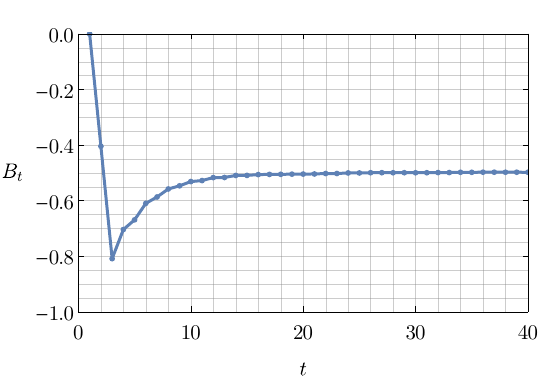}
\caption{Numerically computed coefficient $B_t$ up to $t=40$, for the coin angle $\theta=\frac{2\pi}{5}$. We observe that $B_t$ flattens and barely changes after $t\approx 20$ steps. For other values of $\theta$ $B_t$ shows similar behavior.}
\label{fig:Rn-numerical-conv}
\end{figure}

\begin{figure}[t]
\includegraphics[width=\linewidth]{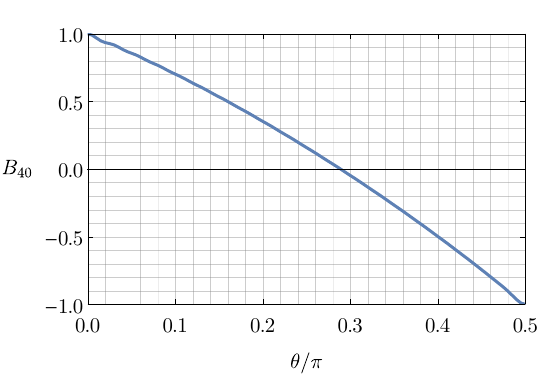}
\caption{Numerically computed coefficient $B_{40}$ from Eq.~(\ref{eq:P-p-expansion}) against the coin angle $\theta$. We see that $B_{40}$ is negative for $\theta \gtrsim 0.289 \pi $, showing that in this range of coins the return probability is a decreasing function for small values of $p$.}
\label{fig:Rn-numerical-all}
\end{figure}

The cross-over point $\theta_*$, where the recurrence probability becomes a decreasing function for small $p$, can be estimated by finding the coin angle $\theta$ where the first derivative of $R_t(p)$ at $p=0$ vanishes for sufficiently large $t$. To evaluate $R_t'(p=0)$ we consider a perturbation expansion of the superoperator $\mathcal{M}$ (\ref{eq:one-step}). We rewrite $\mathcal{M}$ in the form
\begin{equation}
\mathcal{M} = \mathcal{M}_0 + p \mathcal{N},
\end{equation}
where $\mathcal{M}_0$ denotes the unitary case $p=0$, and $\cal N$ acts on the density matrix $\rho$ according to
\begin{eqnarray}
\label{supop:n}
\nonumber \mathcal{N}\hat\rho & = & \hat{\pi}_0' \left( \frac{1}{2}(\hat I_c\otimes\hat T)\ \hat\rho\ (\hat I_c\otimes\hat T^{\dagger})  +{} \right. \\
& & \left. {}+ \frac{1}{2}(\hat I_c\otimes\hat T^{\dagger})\ \hat\rho\ (\hat I_c\otimes\hat T)  - \hat U \hat\rho \hat U^\dagger \right)\hat{\pi}_0'.
\end{eqnarray}
For $p\ll 1$ we neglect events where the classical random walk occurs more than once, resulting in expansion of $t$ iterations to first order in $p$ in the form
\begin{multline}
\label{eq:M-p-expansion}
\mathcal{M}^{\circ t} = \mathcal{M}_0^{\circ t} + p \sum_{k=0}^{t-1}  \mathcal{M}_0^{\circ(t-k-1)}\circ \mathcal{N}\circ  \mathcal{M}_0^{\circ k}  + O(p^2).
\end{multline}
With this formula we expand the finite-time return probability in powers of $p$ as
\begin{equation}
\label{eq:P-p-expansion}
R_t(p) = R_t(p=0) + p B_t+ O(p^2),
\end{equation}
where $R_t(p=0)$ is the probability of return within $t$ steps for the unitary case. The first-order term $B_t$ provides the desired first derivative $R_t'(p=0)$. We find that it can be expressed as a sum
\begin{equation}
\label{bt:sum}
    B_t = \sum_{k=0}^{t-1} B^{(k)}_t
\end{equation}
with
\begin{equation}
\label{btk}
B^{(k)}_t = -\Tr \qty( \mathcal{M}_0^{\circ(t-k-1)} \circ \mathcal{N} \circ \mathcal{M}_0^{\circ k}\hat\rho(0)),
\end{equation}
the minus sign owing to the complementarity of survival and return probabilities.
More explicitly, for a pure initial state $\ket{\Psi(0)}$ (which is without loss of generality) we obtain
\begin{align}
\label{eq:Rk}
B^{(k)}_t = & \left\| (\hat{\pi}_{0}' \hat U)^t \Psi(0) \right\|^2 - \\
\nonumber  & -\frac12 \left\| (\hat{\pi}_{0}' \hat U)^{t-k-1} \hat{\pi}_{0}' (\hat I_c \otimes \hat T)(\hat{\pi}_{0}' \hat U)^k \Psi(0) \right\|^2  - \\
\nonumber &  - \frac12 \left\| (\hat{\pi}_{0}'\hat  U)^{t-k-1} \hat{\pi}_{0}' (\hat I_c \otimes \hat T^\dagger)(\hat{\pi}_{0}' \hat U)^k \Psi(0) \right\|^2.
\end{align}
The decrease of the return probability emerges when the first derivative at zero  $R_t'(p=0) = B_t$ is negative. The sum $B_t$ can be efficiently evaluated numerically; moreover, it saturates with increasing $t$ rather quickly, as illustrated in Fig.~\ref{fig:Rn-numerical-conv}. Fig.~\ref{fig:Rn-numerical-all} shows $B_t$ as a function of $\theta$ for $t=40$, the rightmost point of Fig.~\ref{fig:Rn-numerical-conv}. We observe that $B_{40}$ is negative for $\theta \gtrsim 0.289 \pi$. A more precise estimate of $\theta_*$ using a higher number of steps is presented in Figure~\ref{fig:B100}.

A potential drawback of the above method is that $B_t$ is the initial slope of $R_t(p)$, which is just a convergent to $R(p)$, and the limit of $B_t$ as $t \to \infty$ may not be representative of $R'(0)$ even if both quantities are finite, due to the order of taking the limit and the derivative. This gives an incorrect prediction of the latter value in at least one case: for $\theta = \pi / 2$, there is a visible discrepancy between the numerical value $B_{40} \approx -1$ and the flat character of $R(p)$ in Fig.~\ref{fig:results}, which in this case can be found exactly to be constant 1, see Appendix \ref{app:e} for the derivation. The reason turns out to be that the derivatives of $R_t(p)$ for $\theta = \pi/2$ do not converge uniformly near $p=0$, as the interval where the function is decreasing gradually shortens in $t$ until it disappears completely.

\begin{figure}[t]
\includegraphics[width=\linewidth]{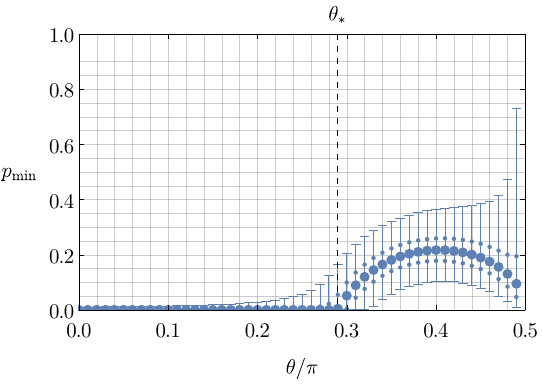}
\caption{Numerically estimated positions of minima of the recurrence probability with varying $p$. The behaviour of $\tilde R_z$ with $z = 0.99999$ was used for the estimate. The minimum starts departing from $p_\text{min}=0$ at $\theta = \theta_* \approx 0.2892\pi$. The uncertainty of $p_\text{min}$ is visualized as the interval where $\tilde R_z$ differs by no more than $0.01$ (fences) or no more than $0.001$ (points) from the identified minimum. Binary search of 15 iterations was employed to find both the $p_\text{min}$ and the crossover points for the uncertainty estimates.}
\label{fig:minima}
\end{figure}

In order to provide an alternative argument for such corner cases, we strengthen our evidence by separately investigating the position $p_{min}$ of minima of $\tilde R_z(p)$. Along with an estimate of deviation of the numerical approximants from the limit value, this negates the possibility that the dip in the recurrence probabilities would flatten out in the infinite step limit. The numerical estimates of $p_{min}$ for different values of the coin angle $\theta$ are presented in Figure~\ref{fig:minima}.

\section{Analytic solution of the case \texorpdfstring{$\theta = \pi/2$}{theta = pi/2}}
\label{app:e}

The case $\theta = \pi/2$ is special in several respects. As the numerical data indicates, the dip in the recurrence probability $R(p)$ that appeared in $\theta > \theta_*$ approaches $p = 0$ in position and $1$ in value. Visually inspecting Fig.~\ref{fig:results}, however, the approximant $\tilde R_z(p)$ displays a slow decline across the entire interval. The negative derivative at $p = 0$, compared with other values of $\theta$, displays inconsistency with the trend predicted by the first-order perturbation analysis. Here we show that this special case can be confirmed analytically to give a flat value $R(p) = 1$, proving that any observed decrease is an effect of finite $t$ or $z \lneq 1$.

For $\theta = \pi/2$ the coin operator \eqref{eq:coinOpr} does not mix the basis states, so overall the quantum walk step becomes a permutation of the $\{\ket{c,x}\}$ basis
\begin{equation}
    \begin{aligned}
        \hat U \ket{R, x} &= \ket{L, x - 1}, \\
        \hat U \ket{L, x} &= \ket{R, x + 1}, \quad \forall x \in \mathbb{Z} \\
    \end{aligned}
\end{equation}
and hence the entire system can be described classically. We define classical states $(c,x)$ in exact correspondence to the chosen basis states. The transition rule for one step takes the form
\begin{equation}
    \begin{aligned}
        (R,x) &\to \begin{cases}
            (L, x-1) & \text{(probability $1-p$)}, \\
            (R, x+1) & \text{(probability $p/2$)}, \\
            (R, x-1) & \text{(probability $p/2$)}, \\
        \end{cases} \\
        (L,x) &\to \begin{cases}
            (R, x+1) & \text{(probability $1-p$)}, \\
            (L, x+1) & \text{(probability $p/2$)}, \\
            (L, x-1) & \text{(probability $p/2$)}. \\
        \end{cases} \\
    \end{aligned}
\end{equation}
In both cases, the three lines correspond to applying the Kraus operators $\hat E_0$, $\hat E_1$, $\hat E_2$ (or: QW step, right shift, left shift), in this order.

The walk can return to the initial position in $n$ steps if and only if $n$ is even. From the initial state $(R, 0)$, it may return to $(R,0)$ again if the number of right ($r$) and left ($l$) steps is the same and the number of QW steps ($q$) even, in any order, for a total probability of
\begin{eqnarray}
\begin{aligned}
    &P((R,0) \to (R,0) | n = 2m) \\
    &= \sum_{r=0}^m \frac{(2m)!}{(2r)!(2m-2r)!} \left(\frac{p}{2}\right)^{2r} (1-p)^{2m-2r}.
\end{aligned}
\end{eqnarray}
It may also return to position 0 with the $L$ coin state for $r = l + 1$ and $q$ odd, again in any order, with total probability
\begin{eqnarray}
        & & P((R,0) \to (L,0) | n = 2m) =  \\
\nonumber    & & = \sum_{l=0}^{m-1} \frac{(2m)!}{(2l+1)!(2m-2l-1)!} \left(\frac{p}{2}\right)^{2l+1} (1-p)^{2m-2l-1}.
    \end{eqnarray}
The return probabilities from $(L,0)$ to $(L,0)$ or $(R,0)$, respectively, are the same two values.

Combining the probabilities of $(R,x)$ and $(L,x)$ in a column vector, the four return probabilities form a matrix $R_n$ acting on the initial state. The respective sums don't seem to allow expression in a simpler form than the equivalent hyper-geometric, but the generating function
\begin{eqnarray}
    \tilde R(z) = \sum_{n=0}^\infty z^n R_n
\end{eqnarray}
can be found as
\begin{eqnarray}
    \tilde R(z) = \frac12 \begin{pmatrix}
        \frac1u + \frac1v & \frac{a}{uc} - \frac{b}{vc} \\
        \frac{a}{uc} - \frac{b}{vc} & \frac1u + \frac1v
    \end{pmatrix}
\end{eqnarray}
with
\begin{equation}
    \begin{aligned}
        a &= 1 - (1-p) z, \\
        b &= 1 + (1-p) z, \quad     c = p z, \\
        u &= \sqrt{a^2 - c^2}, \quad        v = \sqrt{b^2 - c^2}.
    \end{aligned}
\end{equation}
The first-return probabilities are similarly described by a matrix $Q_n$. Its generating function $\tilde Q(z)$ relates to $\tilde R(z)$ through the renewal equation in the form
\begin{equation}
    \tilde Q(z) = I - \tilde R(z)^{-1}.
\end{equation}
The inverse of $\tilde R(z)$ can be written using the symbols introduced earlier as
\begin{equation}
    \tilde R(z)^{-1} = \frac12 \begin{pmatrix}
        a v + b u & c u - c v \\
        c u - c v & a v + b u
    \end{pmatrix}.
\end{equation}
In the limit $z \to 1_-$, all its elements converge in consequence of
\begin{equation}
    \begin{aligned}
        \lim_{z \to 1_-} a &= p, \\
        \lim_{z \to 1_-} b &= 2 - p, \\
        \lim_{z \to 1_-} c &= p, \\
        \lim_{z \to 1_-} u &= 0, \\
        \lim_{z \to 1_-} v &= 2 \sqrt{1 - p}.
    \end{aligned}
\end{equation}
From this the matrix
\begin{equation}
    \lim_{z \to 1_-} \tilde Q(z) = \sum_{n = 0}^{\infty} Q_n
\end{equation}
can easily be found in the form
\begin{equation}
    \begin{pmatrix}
        1 - p \sqrt{1-p} & p \sqrt{1-p} \\
        p \sqrt{1-p} & 1 - p \sqrt{1-p}
    \end{pmatrix}.
\end{equation}
For an initial probability vector $(p_R, p_L)^T$ at $x = 0$, the recurrence probability is the sum of returns to $(R,0)$ and to $(L,0)$, thus
\begin{equation}
    \begin{aligned}
    R &= \begin{pmatrix} 1 & 1 \end{pmatrix} \begin{pmatrix}
        1 - p \sqrt{1-p} & p \sqrt{1-p} \\
        p \sqrt{1-p} & 1 - p \sqrt{1-p}
    \end{pmatrix} \begin{pmatrix} p_R \\ p_L \end{pmatrix} \\
    &= p_R + p_L = 1
    \end{aligned}
\end{equation}
independently of the mixing parameter $p$.

\section{DTQSW interpolating between QW and correlated RW}
\label{app:f}

Let us introduce a model of DTQSW which interpolates between a QW and a correlated RW. First, we review the classical model \cite{patlak_random_1953,weiss_applications_2002,cenac_persistent_2018}, which is less well-known than the balanced RW. However, it is in fact closer to the QW than the balanced RW. Indeed, the similarity between QWs and correlated RWs allowed to apply the methods developed for QWs to prove limit theorems and investigate the absorption problem for correlated RWs \cite{konno_limit_2009}, or study the generating functions of recurrence probabilities for the CRW \cite{kiumi_return_2022}. 

The rules of evolution of correlated RW on a line are such that the particle has a given probability to continue in the same direction or revert, rather than continue left or right. We can describe it in a similar way as the QW, however, instead of probability amplitudes we work directly with probabilities. Denote by $p(x,t) = \qty(p_R(x,t),p_L(x,t))^T$ a two-component vector of probabilities corresponding to the particle being at point $x$ after $t$ steps which was heading right or left in the previous step. The time-evolution of probabilities is then given by
\begin{equation}
    p(x,t+1) = P_R\ p(x-1,t) + P_L\ p(x+1,t),
\end{equation}
where $P_{R,L}$ are the top and bottom rows of the column-stochastic matrix of the correlated random walk
\begin{eqnarray}
\nonumber    P_R & = & \begin{pmatrix}
        p_{RR} & p_{RL} \\
        0 & 0 
    \end{pmatrix}, \\
P_L & = & \begin{pmatrix}
        0 & 0 \\
        p_{LR} & p_{LL} \\
    \end{pmatrix}, \\ \nonumber
    p_{RR} + p_{LR} &=& p_{RL} + p_{LL} = 1.
\end{eqnarray}
Here $p_{uv}$ is the probability that the particle heading in the direction $v=L,R$ in the previous step will move in the direction $u=L,R$. One can see that the correlated RW is related to the coined QW model. Indeed, QW with the coin (\ref{eq:coinOpr}) can be seen as a quantum analogue of the correlated RW where the transition probabilities are given by
\begin{eqnarray}
    \nonumber p_{RR} & = & p_{LL} = \cos^2\theta, \\
    p_{LR} & = & p_{RL} = \sin^2\theta,
\end{eqnarray}
to which it exactly collapses if position is measured after every step, or if the coin undergoes complete phase decoherence before and after the coin toss operation.
We point out that for $\theta=\pi/4$ all transition probabilities $p_{uv}$ are equal to $1/2$, and the correlated RW has the same spatial probability distribution as the usual balanced RW. 

Decohering the coin state before and after the coin toss operation may be described by a CPTP map with four Kraus operators
\begin{equation}
\begin{aligned}
    \hat F_{uv} &= \hat S \left(\ketbra{u}{u}\otimes \hat I_p\right) \hat C \left(\ketbra{v}{v}\otimes \hat I_p\right), \\
    u, v &\in \{R, L\}.
\end{aligned}
\end{equation}
For $u, v = R$, for example, we may write
\begin{equation}
    \begin{aligned}
        \hat F_{RR}
        &= \hat S \left(\ket{R}\mel{R}{\hat C_\theta}{R} \bra{R} \otimes \hat I_p\right) \\
        &= \cos \theta \dyad{R}{R} \otimes \hat T
    \end{aligned}
\end{equation}
(see \eqref{eq:stepOpr}), which in the action of $\hat\rho \mapsto \hat F_{RR} \hat \rho \hat F_{RR}^\dagger$ means an incoherent right shift with probability $\cos^2 \theta = p_{RR}$, projecting the coin onto the $\ket{R}$ state in the process. The other three operators similarly implement the other three classical transitions with the appropriate probabilities $p_{uv}$, leaving the coin state in one of the basis states. In this way, the correlated RW is implemented in the QW state space, adapting the quantum coin state and the existing operators $\hat C$, $\hat S$ for the classical protocol.

Noticing further that the coin projections commute with $\hat S$, this paves the obvious way for interpolating between the two walks by introducing the probability parameter $p \in [0,1]$ and a CPTP map
\begin{align}
{\mathcal C} \hat\rho = \sum_{j=0}^4 \hat F_j \hat\rho \hat F_j^\dagger, 
\label{cptp:qw:crw}
\end{align}
with five Kraus operators,
\begin{equation}
\label{kraus:crw}
    \begin{aligned}
        \hat F_0 &= \sqrt{1-p} \hat U, \\
        \hat F_1 &= \sqrt{p} \left( \dyad{R}{R} \otimes \hat I_p \right) \hat U \left( \dyad{R}{R} \otimes \hat I_p \right), \\
        \hat F_2 &= \sqrt{p} \left( \dyad{R}{R} \otimes \hat I_p \right) \hat U \left( \dyad{L}{L} \otimes \hat I_p \right), \\
        \hat F_3 &= \sqrt{p} \left( \dyad{L}{L} \otimes \hat I_p \right) \hat U \left( \dyad{R}{R} \otimes \hat I_p \right), \\
        \hat F_4 &= \sqrt{p} \left( \dyad{L}{L} \otimes \hat I_p \right) \hat U \left( \dyad{L}{L} \otimes \hat I_p \right).
    \end{aligned}
\end{equation}

We point out that the CPTP maps ${\mathcal C}$ and ${\mathcal T}$ (\ref{eq:teDm}) are different even for $\theta=\pi/4$, where the correlated random walk results in the same distribution as the balanced random walk. The reason is that the balanced random walk does not utilize the coin states at all. In terms of the Kraus operators, $\hat E_{1,2}$ describing the balanced random walk in (\ref{eq:teDm2}) act independently of the coin state, whereas $\hat F_j$ include projections onto $\ket{R}$ or $\ket{L}$. 

\begin{figure}[t!]
\includegraphics[width=\linewidth]{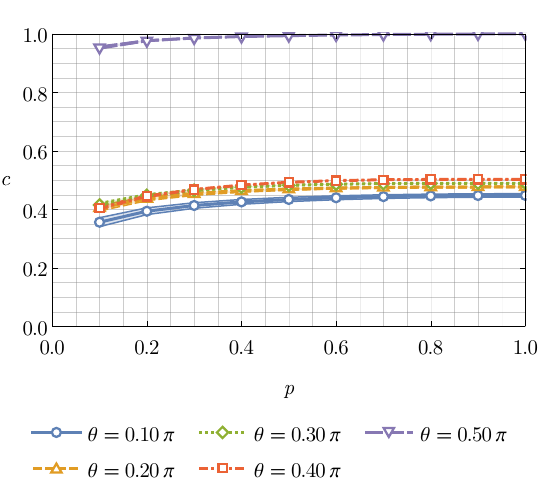}
\caption{Fit results of $\tilde{R}_z(p)$ for varying $z$ parameter for the DTQSW interpolating between QW and a correlated RW. The fitting function is
$1 - b(1 - z)^c$. For $\theta = 0$ or $p = 0$ the limit value is not 1, so the fit is ill-conditioned (omitted from the plot). For all other values the validity of the fit supports the hypothesis that the limit is equal to 1. The best fit value and standard error of the exponent $c$ are plotted for various combinations of $\theta$ and $p$. The same values for $z$ and $N_\text{max}$ were used as for Figure~\ref{fig:convergence-z-in-p}.}
\label{fig:fit:crw}
\end{figure}

We have evaluated the recurrence probability $R_t(p)$ by directly simulating the monitored evolution of the CPTP map $\mathcal{C}$, and also calculated the approximation of the recurrence probability $\tilde{R}_z(p)$  using the generating functions. The data presented in Figure~\ref{fig:recur:crw} indicates that the recurrence probability is a step function equal 1 for $p\neq 0$. This hypothesis is supported by fitting the data obtained from the generating function on $1 - b(1-z)^c$, see Figure~\ref{fig:fit:crw}.

\begin{figure}
\includegraphics[width=\linewidth]{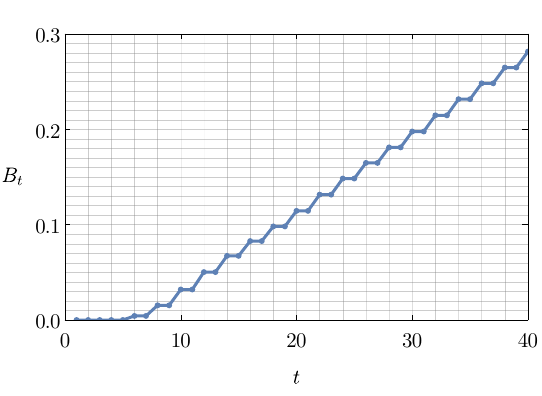}
\caption{First-order expansion of the return probability $R_t(p)$ for the DTQSW interpolating between QW and correlated RW. The coin angle is chosen as $\theta = 2\pi/5$. Unlike for the interpolation between QW and balanced RW (see Figure~\ref{fig:Rn-numerical-conv}), $B_t$ does not saturate. Rather, it increases linearly with $t$. The same behavior applies to all values of $\theta\neq 0$ for large $t$.}
\label{fig:1st:crw}
\end{figure}

We also consider first order expansion (\ref{eq:P-p-expansion}) of the exact return probability $R_t(p)$ in $p$, which in the case of DTQSW interpolating between the QW and the correlated RW is given by a sum (\ref{bt:sum}) of terms
\begin{align}
\label{eq:Bk2} B^{(k)}_t = & \left\| (\hat{\pi}_{0}' \hat U)^t \Psi(0) \right\|^2 - \\
\nonumber & - \left\| (\hat{\pi}_{0}' \hat U)^{t-k-1} \hat{\pi}_{0}' \hat F_{RR} (\hat{\pi}_{0}' \hat U)^k \Psi(0) \right\|^2 - \\
\nonumber & - \left\| (\hat{\pi}_{0}' \hat U)^{t-k-1} \hat{\pi}_{0}' \hat F_{RL} (\hat{\pi}_{0}' \hat U)^k \Psi(0) \right\|^2 - \\
\nonumber & - \left\| (\hat{\pi}_{0}' \hat U)^{t-k-1} \hat{\pi}_{0}' \hat F_{LR} (\hat{\pi}_{0}' \hat U)^k \Psi(0) \right\|^2 - \\
\nonumber  & - \left\| (\hat{\pi}_{0}'\hat  U)^{t-k-1} \hat{\pi}_{0}' \hat F_{LL} (\hat{\pi}_{0}' \hat U)^k \Psi(0) \right\|^2 .
\end{align}
The results are illustrated for $\theta = \frac{2\pi}{5}$ in Figure~\ref{fig:1st:crw} where we plot $B_t$ as a function of the number of steps $t$. We observe that $B_t$ approaches a linear increase, in contrast to the case of the map $\cal T$ where $B_t$ saturates, see Figure~\ref{fig:Rn-numerical-conv}. The same applies to all values of $\theta \neq 0$. Hence, the first derivative $R'_t(0) = B_t$ at $p=0$ increases with $t$, consistent with the result of the generating function given by $\tilde{R}_z(p)$, which approaches a step function.

\end{document}